\documentstyle[pra,epsf,aps,twocolumn]{revtex}

\def\qn{x}
\def\phin{q}
\def\lfrac#1#2{#1/#2}
\def\db{\,\, {\bar{} \!\!d}\!\,\hspace{0,5pt}}
\def\comment#1{}
\def\lfrac#1#2{#1/#2}

\begin{document}
\sloppy
\title{Coordinate Independence of  of Quantum-Mechanical Path Integrals
}

\author{H.~Kleinert\thanks{E-mail: kleinert@physik.fu-berlin.de} and
     A.~Chervyakov\thanks{On leave from LCTA, JINR, Dubna, Russia
                   E-mail: chervyak@physik.fu-berlin.de}
                      \\ Freie Universit\"at Berlin\\
          Institut f\"ur Theoretische Physik\\
          Arnimallee14, D-14195 Berlin}
\maketitle
\begin{abstract}
We develop simple rules for
performing  integrals  over products of
distributions
in coordinate space.
Such products occur
in
perturbation expansions
of path integrals in curvilinear coordinates,
where the interactions contain terms of the form
$\dot q^2q^n$, which give rise to
highly singular Feynman integrals.
The new rules ensure
the invariance
of perturbatively defined path integrals
under coordinate transformations.
\end{abstract}
\section{Introduction}
In the previous papers \cite{1,2},
we have presented
a diagrammatic proof of reparametrization
invariance of perturbatively defined
quantum-mechanical path integrals.
%, if the Feynman diagrams
%are calculated by dimensional regularization in momentum space.
The proper perturbative definition of path integrals
was shown to
require an extension
to a
functional integral in $D$ spacetime,
and a subsequent
analytic continuation to $D=1$.
In Ref.~\cite{1}
the perturbative calculations
were performed in momentum space,
where
Feynman integrals
in
a continuous number of dimensions
$D$ are known from the prescriptions of
't~Hooft and M.~Veltman\cite{3}.
In Ref.~\cite{2} we have found
the same results directly
from the
Feynman integrals in the $1-\varepsilon$-dimensional
time space
with the help
of the Bessel representation of Green
functions.
The coordinate space calculation
is interesting
for many applications,
for instance, if one wants to obtain the effective
action of a field system in curvilinear
coordinates, where the kinetic term depends
on the dynamic variable.
Then one
needs rules for performing temporal integrals
over
Wick contractions
of local fields.

In this note we want to show
that the  reparametrization
invariance of perturbatively defined
quantum-mechanical path integrals
can be obtained in the coordinate space
with the help
of a simple
but quite general arguments
based on the inhomogeneous field equation
for the Green function,  and
rules of the partial integration.
The prove does not require the
calculation of the Feynman
integrals separately and
remains valid for the functional
integrals in an arbitrary space-time
dimension $D$.

\section{Problem with Coordinate Transformations}
Recall the origin of the difficulties
with coordinate transformations
in path integrals.
Let $\qn(\tau )$ be the euclidean coordinates of a quantum-mechanical
point particle of unit mass in a harmonic potential  $\omega^2\qn^2/2$
as a function of the imaginary time $\tau =-it$.
Under a coordinate transformation
 $\qn(\tau)\rightarrow \phin(\tau )$
defined by $\qn (\tau )=f(\phin(\tau ))=
\phin(\tau )+\sum _{n=2}^\infty a_n
\phin^n(\tau )$,
the kinetic term $\dot\qn^2(\tau )/2$
goes over into $\dot\phin^2(\tau )f'{}^2(\phin(\tau ))/2$.
If the path integral over $\phin(\tau )$ is performed
perturbatively, the
expansion terms
contains temporal integrals
over
Wick
contractions which,
after suitable partial integrations,
are products of the following basic correlation functions
\begin{eqnarray}
\Delta (\tau -\tau ')&\equiv& \langle \phin(\tau )\phin(\tau ')\rangle=
\hspace{0mm}\raisebox{-1mm}{\mbox{\input 1.tps }} ,~\label{@D1}\\
\partial _\tau \Delta(\tau -\tau ')&\equiv &\langle\dot\phin(\tau )\phin(\tau ')\rangle
=\hspace{0mm}\raisebox{-1mm}{\mbox{\input 3.tps }} ,~\label{@D2}\\
\partial_ \tau \partial_{\tau '}\Delta (\tau-\tau ')&\equiv& \langle\dot\phin(\tau )
\dot\phin(\tau ')\rangle
=\hspace{0mm}\raisebox{-1mm}{\mbox{\input 2.tps }}.~\label{@D3}
\label{@}\end{eqnarray}
The right-hand sides define the line symbols
to be used in Feynman diagrams for the interaction terms.

Explicitly,
the first correlation function reads
\begin{equation}
 \Delta (\tau ,\tau ')=\frac{1}{2\omega}e^{-\omega|\tau -\tau '|}.
\label{@del1}\end{equation}
The second correlation function (\ref{@D2})
has a discontinuity
\begin{equation}
\partial _\tau \Delta(\tau ,\tau ') =
     - \frac{1}{2} \epsilon (\tau - \tau ') e^{-\omega|\tau -\tau '|} ,
\label{@del2}\end{equation}
where
\begin{equation}
\epsilon (\tau - \tau ')\equiv 2\int_{-\infty}^\tau  d\tau''  \delta (\tau'' -\tau ')
\label{@}\end{equation}
is a distribution
which has a
jump at $\tau =\tau '$.
The third correlation function (\ref{@D3}) contains a
$ \delta $-function:
\begin{equation}
 \partial_ \tau \partial_{\tau '}\Delta (\tau, \tau ') =
  \delta(\tau -\tau ') - \frac{\omega}{2}e^{-\omega|\tau -\tau '|} ,
\label{@del3}\end{equation}
The temporal
integrals in $\tau $-space over products of such distributions
are undefined.

In our previous papers \cite{1,2}
we have shown
that a
unique perturbation expansions leading
to a
reparametrization invariant theory
is obtained by extending the path integral to a $D$-dimensional
functional integral, and by performing the
perturbation
expansion in $D$-dimensional space,
with a limit $D\rightarrow 1$ taken at the end.

In this note we
shall set up simple rules for
integrals over products
of the correlation functions
in the extended coordinate space
the same results.

\section{Model System}
To be specific, we shall prove the
coordinate independence
of the
exactly solvable
path integral
of a point particle of unit mass in a harmonic
potential $\omega^2 \qn ^2/2$,
over a large imaginary-time interval $ \beta $,
\begin{equation}
  Z_ \omega  = \int  {\cal D} \qn  (\tau)\,
e^{-{\cal A}_{ \omega } [\qn ]}
= e^{- \rm Tr \log (-\partial^2 + \omega^2)} =
  e^{-\beta\lfrac{\omega}{2}}.
\label{m1}\end{equation}
The action is
\begin{equation}
{\cal A}_ \omega   = \frac{1}{2}\,\int\,
d \tau \left[\dot \qn ^2(\tau ) + \omega^2 \qn ^2(\tau )
\right].
\label{m2}\end{equation}
 A coordinate transformation
turns  (\ref{m1}) into a path integral
with a
singular perturbation expansion.

For simplicity we assume the coordinate transformation
to preserve the symmetry $\qn \leftrightarrow -\qn $ of the initial oscillator,
such its power series expansion starts out like
$\qn  (\tau)=f(\phin (\tau)) = \phin  - {g}\phin ^3/3 + {g^2}a\phin ^5/5 - \cdots~$,
where $g$ is a smallness parameter, and $a$ an extra parameter.
We shall see that the perturbation expansion is independent of $a$, such that $a$
will merely
serve
to check the calculations.
The transformation changes the partition function
(\ref{m1})
into
\begin{equation}
  Z = \int  {\cal D} \phin  (\tau)\,
e^{-{\cal A}_{J} [\phin ]}
e^{-{\cal A} [\phin ]},
\label{m3}\end{equation}
where the transformed action
$ {\cal A}_{} [\phin ] =
{\cal A}_{ \omega } [\phin ] +
{\cal A}_{\rm int} [\phin ]
$
is decomposed into a free
part
\begin{equation}
{\cal A}_{ \omega } [\phin ]= \frac{1}{2}\,\int\, d \tau [\dot \phin ^2(\tau )
+\omega^2 \phin ^2 (\tau)] ,
\label{m4}\end{equation}
and an interacting part, which reads to
second order in  $g$:
\begin{eqnarray}
&&\!\!~
{\cal A}_{\rm int} [\phin ] = \frac{1}{2} \int d \tau
\left\{-g\left[ 2\dot \phin ^2  (\tau) \phin ^2 (\tau)
+ \frac{2\omega^2}{3}\phin ^4 (\tau) \right]\right.
\nonumber\\&&\left.  ~
 + g^2
\left[\left(1 + 2a\right) \dot \phin ^2 (\tau)  \phin ^4 (\tau)
+ \omega^2 \left(\frac{1}{9} + \frac{2a}{5}\right)
\phin ^6  (\tau)\right]\right\} . \label{m5}
%\\&& \quad\quad\quad +\, {\rm O}(g^3)\,.
\end{eqnarray}
The
exponent in (\ref{m3})
contains an additional
effective action
$ {\cal A}_J[\phin ]$
coming from the
Jacobian of the coordinate transformation:
\begin{equation}
 {\cal A}_J[\phin ]=
-\delta (0)\int d \tau\,\log \,\frac{\delta f(\phin (\tau))}{\delta \phin (\tau)}.
\label{m6}\end{equation}
This has the power series expansion
\begin{eqnarray}
 &&
\, {\cal A}_J[\phin ]=
-\delta (0)\int d \tau\left[-g \phin ^2(\tau) +
g^2 \left(a - \frac{1}{2}\right) \phin ^4 (\tau)
\right] . \label{pix@m8}
%\\&&\quad\quad\quad + \, {\rm O}(g^3)~~~~~~~~~~~~~~~.
\end{eqnarray}

For $g=0$,
the transformed partition function
(\ref{m3}) coincides with (\ref{m1}).
When expanding   $Z$ of Eq.~(\ref{m3})
in powers of $g$,
we obtain
a sum of Wick contractions
with associated  Feynman diagrams contributing to each order $g^n$.
This sum must vanish
to ensure coordinate invariance of the path integral.

By considering only
connected Feynman diagrams,
we shall obtain an expansion for the free energy
\begin{equation}
F=F_ \omega +\sum _{n=1}g^nF_n,
\label{m7}\end{equation}
where $F_{\omega}$
is the free energy of the unperturbed
harmonic oscillator
(\ref{m1}).
The
coordinate invariance is ensured by the vanishing
of all expansion terms $F_n$.

\section{Expansion Terms of Free Energy Density}
The graphical expansion for
the ground state energy
will be carried here only up to three loops.
The diagrams are composed of the
three
types of lines in (\ref{@D1})--(\ref{@D3}), and
new interaction vertices
for each
power of $g$.
The diagrams coming from the Jacobian action
(\ref{pix@m8}) are easily recognized by an accompanying power of $ \delta (0)$.

To lowest order in $g$, there
exists only three diagrams, two
originated from the
interaction (\ref{m5}), one
from the Jacobian action (\ref{pix@m8}):
\begin{equation}
\!\!\!\!\!\!\!\!\!\!\!-F_1=
-\,g\bigg[\,\hspace{0mm}\raisebox{-1mm}{\mbox{\input 6.tps }} +\,\omega^{2}\hspace{-27mm}\raisebox{-11.57mm}{\mbox{\input inf.tps }} ~~~~~~~~~~~~~~
~~~
~~~
~~~
 -\,\delta (0) \hspace{0mm}\raisebox{-1mm}{\mbox{\input 0dot.tps }}
\bigg].
\label{f1}\end{equation}
~\\[-1.2cm]

To order $g^2$,
we distinguish several contributions.
First there are two three-loop local diagrams
coming
from the interaction (\ref{m5}),
and one two-loop local diagram
from the Jacobian action (\ref{pix@m8}):
\begin{eqnarray}
 &&\!\!\!\!\!\!\!-F_2^{(1)}=  g^2\,\Bigg[\,
 ~\,3 \left(\frac{1}{2} + a\right)\!\!\hspace{0mm}\raisebox{-3.2mm}{\mbox{\input 7.tps }} +
15\, \omega^{2} \left(\frac{1}{18} + \frac{a}{5}\right)
 \hspace{-27mm}\raisebox{-13.7mm}{\mbox{\input clover.tps }}~~~~~~~~~~~~~~~~~~
~\nonumber \\[-1.2cm] \nonumber\\&&~
~~~~~~~~~~~~- 3\left(a - \frac{1}{2}\right)\, \delta (0)\!
\hspace{-27mm}\raisebox{-11.57mm}{\mbox{\input inf.tps }} ~~~~~~~~~~~~~~
~~~~~~
\quad\Bigg]\,.  \nonumber \\[-.9cm]&&
\label{f2}\end{eqnarray}
We call a diagram local if it involves
no temporal time integral.
The Jacobian
action (\ref{pix@m8}) contributes further
the nonlocal diagrams:
\begin{eqnarray}
 &&~~~~~\!\!\!\!\!\!\!\!\!\!\!\!-F_2^{(2)}=
-\frac{g^2}{2!}\Big\{
 2\delta^2 (0) \!\!\hspace{0mm}\raisebox{-1mm}{\mbox{\input 0dotdot.tps }}
\nonumber \\&& ~~~~~~~~~~~~\!
 \!\!- 4\delta (0) \Big[\!\!
\hspace{0mm}\raisebox{-1mm}{\mbox{\input 6dot.tps }}\!\!\!\!\!   ~~+  \!\!\!
\hspace{0mm}\raisebox{-1mm}{\mbox{\input 6pdot.tps }}
+ 2\,\omega^{2}\!\!\hspace{0mm}\raisebox{-1mm}{\mbox{\input infdot.tps }}\Big]\Big\}.
\label{pix@f3}\end{eqnarray}
In the perturbative calculations
to follow,
we shall use
dimensional regularization where $ \delta (0)=0$,
according to a basic rule
of t'Hooft and Veltman \cite{3}. As a consequence,
the last terms in $F_1$,  $F_2^{(1)}$, and
the entire
$F_2^{(2)}$ are zero. In fact,
the  term $ {\cal A}_J[\phin ]$ may be omitted completely from
the path integral
(\ref{m3}).

The remaining diagrams
are either of the
three-bubble type, or of the watermelon
type, each with all possible combinations
of the three line types (\ref{@D1})--(\ref{@D3}).
The former are
\begin{eqnarray}
&&
-F_2^{(3)}
= -\frac{g^2}{2!}\big[ 4\hspace{0mm}\raisebox{-1.2mm}{\mbox{\input 8.tps }}
\!+\,\,2~\hspace{0mm}\raisebox{-1.2mm}{\mbox{\input 9.tps }}
~+\,\,\,2\,\,\,\hspace{0mm}\raisebox{-1.2mm}{\mbox{\input 10.tps }}
\nonumber \\[1mm]
&&~~~+
 ~8\,\omega^2\!
 \hspace{-27.0mm}\raisebox{-11.5mm}{\mbox{\input threeb1.tps }}~~~
~~~~~~~~~~~~~~~~~~~
+ 8\omega^2 \!\!\hspace{-27.0mm}\raisebox{-11.5mm}{\mbox{\input threeb2.tps }}~~~~~~~~~~~~~~~~~~~~~~~
 + 8 \omega^4 \hspace{-27.0mm}\raisebox{-11.5mm}{\mbox{\input threeb.tps }}~~~~~
\quad\quad\quad\quad\quad \quad \big] ~,
\label{f4}\end{eqnarray}
~\\[-1.2cm]
and the latter:
\begin{eqnarray}
& & \!\!\!\!\!-F_2^{(4)}\!=
{\!\!-\frac{g^2}{2!}\, 4 \,\!\bigg[\!\!
\!\!\!\hspace{0mm}\raisebox{-2mm}{\mbox{\input 11.tps }}
\!\!\!+ 4\!\!\!\hspace{0mm}\raisebox{-1.95mm}{\mbox{\input 12.tps }}
\!\!+\!\!\!\!\! \hspace{0mm}\raisebox{-1.9mm}{\mbox{\input 13.tps }}
\!\!\!+ \! 4\omega^2\!\! \hspace{-27.0mm}\raisebox{-12.3mm}{\mbox{\input waterm2.tps }}
~~~~~~~~~~~
~~~~~~~~~~~
\! \!+\! \!\frac{2}{3}\omega^4 \!\!\!
\hspace{-27.0mm}\raisebox{-12.3mm}{\mbox{\input waterm.tps }}
~~~~~~~~~~~
~~~~~~~~~~~
\!\bigg].   \!\!\!\!\! }\nonumber\\[-8mm]
\label{f5}
\end{eqnarray}
%~\\[-1.2cm]
Since the equal-time
expectation value $\langle\dot\phin (\tau)\,\phin (\tau)\rangle$
vanishes by Eq.~(\ref{@del2}),
diagrams with a local contraction of a mixed line
(\ref{@D2}) are trivially zero, and
have been omitted.

In our
previous papers\cite{1,2},
all
integrals were calculated individually
in
$D=1-\varepsilon$ dimensions,
taking the limit
$\varepsilon\rightarrow 0$ at the end.
Here we set up
simple rules for
finding the same results, which make
the sum of all
Feynman diagrams
contributing to each order $g^n$ vanish.

\section{Basic Properties of Dimensionally Regularized Distributions}

The path
integral (\ref{m3}) is extended
to an associated functional integral
in a $D$-dimensional coordinate space $x$,
with coordinates $x_\mu\equiv (\tau ,x_2,x_3,\dots)$,
by replacing $\dot\phin^2(\tau )$ in the kinetic term
by $(\partial_\mu\phin(x))^2$,
where $\partial _\mu=\partial /\partial x_\mu$.
 The Jacobian action term
(\ref{m6}) is omitted
in dimensional regularization
because
of Veltman's rule
\cite{3}:
\begin{equation}
 \delta ^{(D)}(0)= \int \frac{d^D k}{(2\pi)^D} =0.
\label{b1}\end{equation}
In our calculations, we shall encounter
generalized $ \delta $-functions, which are multiple derivatives of
the ordinary $ \delta $-function:
\begin{eqnarray}
  \delta ^{(D)} _{\mu_1 \dots \mu_n} (x)& \equiv&
\partial_{\mu_1 \dots \mu_n}  \delta ^{(D)}  (x)\nonumber \\
& =&
\int \db^D k
(ik)_{\mu_1} \dots (ik)_{ \mu_n} e^{ikx},
\label{10}\end{eqnarray}
with
$\partial_{\mu_1 \dots \mu_n} \equiv
\partial_{\mu_1} \dots \partial _{\mu_n}$,
and
with $\db^D k \equiv d^D k \big/ (2 \pi )^D$.
In dimensional regularization,
alle these vanish at the origin
as well:
\begin{equation}
   \delta  _{{\mu_1 }\dots \mu_n}^{(D)} (0) = \int \db^D k (i k)_{\mu_1}
 \dots (i k)_{\mu_n} = 0 ,
\label{11}\end{equation}
which is a more general way of expressing Veltman's rule.
In the extended coordinate space, the correlation function
(\ref{@D1})
becomes
\begin{equation}
 \Delta (x) = \int \frac{d^Dk}{(2 \pi )^D} \frac{e^{ikx}} {k^2 + \omega^2}
,
\label{1}\end{equation}
At the origin,
it has the value
\begin{equation}
  \Delta (0) = \int \frac{\db^D k}{k^2 + \omega^2}  =
    \frac{\omega^{D - 2}}{(4 \pi ) ^{D/2}}  \Gamma \left(1 -
    \frac{D}{2}\right)
 \mathop{=}_{D=1} \frac{1}{2\omega}.
\label{12}\end{equation}
The
 extension of the time derivative (\ref{@D2}),
\begin{equation}
 \Delta _\mu (x) = \int \db^D k \frac{ik_\mu}{k^2 + \omega^2}
   e^{ikx}
\label{13}\end{equation}
vanishes at the origin, $ \Delta _\mu (0) = 0$.
This follows directly from a Taylor series expansion of $1/(k^2 + \omega^2)$
in powers of $k^2$, together with Eq.~(\ref{11}).

The second derivative of $ \Delta (x)$ has the Fourier representation
\begin{equation}
 \Delta _{\mu \nu } (x) = -\,\int \db^D k\, \frac{k_\mu k_\nu}{k^2 + \omega^2}
 e ^{ikx}.
\label{14}\end{equation}
Contracting the indices yields
\begin{equation}
 \Delta_{\mu\mu } (x) = -\,\int \db^D k\, \frac{k^2}{k^2+ \omega^2}
e^{ikx} = -  \delta ^{(D)} (x) + \omega^2  \Delta (x)\,,
\label{15}\end{equation}
which follows from the definition of the correlation function
by the inhomogeneous
field equation
\begin{equation}
(-\partial _\mu ^2+\omega^2)q (x)= \delta^{(D)} (x) .
\label{@eom}\end{equation}
From (\ref{15}) we have
the relation between integrals
\begin{equation}
 \int d^D x\,  \Delta _{\mu \mu } (x) =-1+ \omega^2 \int d^D x\,  \Delta (x),
\label{17}\end{equation}
Inserting Veltman's rule (\ref{b1}) into (\ref{15}), we obtain
\begin{equation}
  \Delta _{\mu\mu } (0) = \omega^2 \,\Delta (0) \mathop{=}_{D=1}
\frac{\omega}{2}.
\label{16}\end{equation}
This ensures the vanishing
of the first-order contribution
(\ref{f1}) to the free energy
\begin{equation}
-F_1 = -g\,\left[
- \Delta _{\mu\mu } (0) + \omega^2  \Delta (0)
 \,\right]\,\Delta (0) = 0 .
\label{b2}\end{equation}

The same equation (\ref{15}) allows us to calculate immediately
the second-order contribution   (\ref{f2})
from the local diagrams
\begin{eqnarray}
&&-F_2^{(1)} = -3g^2 \,
 \left[ \left(\frac{1}{2} + a \right)\Delta _{\mu\mu }(0)
-5 \left(\frac{1}{18} + \frac{a}{5}\right)
\omega^2 \Delta (0)\right]
\nonumber\\
&&\quad\quad\quad\quad\times\,
\Delta^2 (0)
 = - \frac{2}{3} \,\omega^2 \Delta^3 (0)
\mathop{=}_{D\rightarrow 1}  - \frac{1}{12\omega}.
\label{b3}\end{eqnarray}
The other contributions to the free energy in the expansion (\ref{m7}) require
rules for calculating products of two and four distributions, which we are now going
to develop.

\section{Integrals over Products of Two Distributions}
\label{VA}
The simplest integrals of this type are
\begin{eqnarray}
 && \!\!\!\!\!\!\int d^D x\,  \Delta ^2 (x) =  \int \db^D p\, \db^D k\,
   \frac{\delta ^{(D)}(k+p)}{ (p^2 + \omega^2) (k^2 + \omega^2)} \label{s2} \\
 &\! = &\!\!\int \frac{\db^D k\,}{(k^2 + \omega^2)^2} \!=\!
    \frac{\omega^{D - 4}}{(4 \pi ) ^{D/2}}  \Gamma \left(2 \!-\!
    \frac{D}{2}\right)\! =\! \frac{(2-D)}{2\omega^2}\,\Delta (0),
\nonumber \end{eqnarray}
and
\begin{eqnarray}
  & & \int d^D x\,  \Delta ^2 _\mu(x) = - \int d^D x\,  \Delta  (x) \left[
    -  \delta ^{(D)} (x) + \omega^2  \Delta (x) \right]
 \nonumber \\
 & &~~~~~~ =   \Delta (0) - \omega^2 \int d^D x\,  \Delta ^2 (x)
\!=\!
     \frac{D}{2}\,\Delta (0)\,.
\label{rest2-18}\end{eqnarray}
To obtain the second result we have perfomed
a
partial integration and used
(\ref{15}).

In contrast to the integrals (\ref{s2}) and (\ref{rest2-18}), the integral
\begin{eqnarray}
&& \!\!\!\!\!\!\!\!\!\! \!\!\!\!\!\!\!\!\!\! \int d^D x\,  \Delta ^2_{\mu \nu } (x) =
\int \db^D p\,
\db^D k\, \frac{(kp)^2\, \delta ^{(D)}(k+p)}{ (k^2 + \omega^2)(p^2 + \omega^2)}
\nonumber \\
 & = & \int \db^D k\, \frac{(k^2)^2 }{(k^2 + \omega^2)^2} =
 \int d^D x\,  \Delta _{\mu\mu }^2 (x)
\label{19}\end{eqnarray}
diverges formally in $D=1$ dimension.
In dimensional regularization, however,
we may decompose
 $(k^2)^2 = (k^2 + \omega^2)^2 - 2\omega^2 (k^2 + \omega^2) + \omega^4$,
and use (\ref{11}) to evaluate further
\begin{eqnarray}
 & & \int d^D x\,  \Delta ^2_{\mu\mu } (x)\! =\!\! \int \db^D k
   \frac{(k^2)^2 }{(k^2 + \omega^2)^2} \!=\! - 2\omega^2 \int\!
 \frac{\db ^D k\,}{(k^2 + \omega^2)}\nonumber \\
 &&~+\, \omega^4 \!\int\! \frac{\db^D k\,}{(k^2 + \omega^2)^2} \!=\! - 2\omega^2  \Delta
(0)\! + \omega^4 \!\int d^D  x  \Delta ^2 (x).
 \label{20}\end{eqnarray}
 Together with (\ref{s2}), we obtain the finite integrals
\begin{eqnarray}
 && \int d^D x\,  \Delta ^2_{\mu \nu } (x) =
  \int d^D x\,  \Delta ^2_{\mu\mu }(x) =  -2\omega^2  \Delta (0) \nonumber \\
  &&~~~ +\, \omega^4 \int d^D x\,  \Delta ^2 (x) = -
  \left(1+\lfrac{D}{2}\right) \omega^2 \Delta (0)\,.
\label{21}\end{eqnarray}
     An alternative  way of deriving
the equality
(\ref{19}) is
to use partial integrations and the
identity
\begin{equation}
 \partial_\mu  \Delta _{\mu \nu } (x) = \partial_ \nu   \Delta _{\mu\mu }(x),
\label{28}\end{equation}
which follows directly from the Fourier representation
(\ref{13}).
%A partial integration yields directly
%\begin{eqnarray}
%&& \int d^D x\,  \Delta ^2_{\mu\nu} (x)= -
% \int d^D x\,  \Delta_\nu (x) \partial_\mu \Delta_{\mu \nu} (x) \nonumber \\
%&&=
%- \int d^D x\,  \Delta_\nu (x) \partial_\nu \Delta_{\mu \mu} (x)
%%\nonumber \\&=&
%=
% \int d^D x\,  \Delta_{\nu\nu }(x)  \Delta_{\mu \mu} (x) .
%\label{s4a}\end{eqnarray}
%
%
%Eq.~(\ref{21})
%can also be found from partial integrations
%and the inhomogeneous field equation
%(\ref{15}).

Finally, from Eqs.~(\ref{s2}), (\ref{rest2-18}), and (\ref{21}),
we observe the useful identity
\begin{equation}
 \int d^D x\,\left[\Delta ^2_{\mu\nu} (x) +2\omega^2\,
 \Delta^2 _\mu (x)
 + \omega^4 \,\Delta ^2 (x)\right]\,=\,0\,,
\label{ns5}\end{equation}
which together with the  inhomogeneous field equation (\ref{15})
reduces the calculation of the second-order
contribution of all three-bubble diagrams (\ref{f4})
to zero:
\begin{eqnarray}
&&-F_2^{(3)} = -g^2 \Delta^2 (0)
 \nonumber\\
&&
 \times
 \int d^D x\,\left[\Delta ^2_{\mu\nu} (x) +2\omega^2\,
 \Delta^2 _\mu (x)
 + \omega^4 \,\Delta ^2 (x)\right]\,=\,0\,.
\label{ns6}\end{eqnarray}

\section{Integrals Products of Four Distributions}
\label{VB}
More delicate integrals arise
from the watermelon diagrams in (\ref{f5})
which contain
products of four distributions, a nontrivial
 tensorial
structure, and
overlapping divergences
\cite{1,2}.
Consider  the first three
diagrams:
\begin{eqnarray}
\!\!\!\!  \hspace{2mm}\raisebox{-2mm}{\mbox{\input 11.tps }} &=&
\int d^D x\,  \Delta ^2 (x)  \Delta ^2_{\mu \nu }(x).
\label{29f}\\
\!\!\!\!\!\! 4\!\!\!\!\hspace{2mm}\raisebox{-1.9mm}{\mbox{\input 12.tps }} \!~
 &=&~ \!\!4\int\! d^Dx \Delta (x)
 \Delta _\mu(x) \Delta _ \nu (x) \Delta _{\mu \nu }(x),
\label{29ff}\\
~~\!\!\!\!\!\!\!\!\!\! \hspace{2mm}\raisebox{-2mm}{\mbox{\input 13.tps }}  &=&
\int d^Dx\Delta _\mu(x)
\Delta _\mu(x)
\Delta _\nu(x)
\Delta _\nu(x),
\label{29}
\end{eqnarray}
To exhibit the subtleties with the tensorial
structure,
we introduce the integral
\begin{eqnarray}
 I_{D} = \int d^D x\,\Delta ^2 (x)
 \left[ \Delta ^2 _{\mu\nu }(x) -
  \Delta ^2 _{\mu\mu }(x)
 \right] .
\label{30}
\end{eqnarray}
In $D=1$ dimension, the bracket vanishes formally,
but the limit $D\rightarrow1$ of the integral is nevertheless finite.
 We now decompose
the Feynman diagram (\ref{29f}),
into the sum
\begin{eqnarray}
 \int\! d^D x\,  \Delta ^2 (x) \Delta ^2 _{\mu\nu }(x)
 =  \int\! d^D x\,  \Delta ^2 (x) \Delta ^2 _{\mu\mu }(x)
 + I_{D}\,.
\label{29a}\end{eqnarray}

To obtain an analogous
decompositions for the
other two
diagrams (\ref{29ff}) and (\ref{29})
we derive a few useful relations
using
the  inhomogeneous field equation (\ref{15}), partial integrations, and
Veltman's rule (\ref{11}).
First there is
the relation
\begin{equation}
\!\!- \int d^D x\,  \Delta _{\mu\mu }(x)  \Delta ^3(x) =
  \Delta ^3 (0) - \omega^2 \int d^D x\,  \Delta ^4 (x).
\label{31}\end{equation}
By a partial integration, the left-hand side becomes
\begin{equation}
 \int d^D x\,  \Delta _{\mu\mu }(x)  \Delta ^3(x)
=-
3 \int d^D x\,  \Delta ^2_\mu(x)  \Delta ^2(x) ,
\label{nolabel}\end{equation}
leading to
\begin{equation}
\int d^D x\,  \Delta ^2_\mu(x)  \Delta ^2(x) =
\frac{1}{3}\Delta ^3(0) - \frac{1}{3} \omega^2 \int d^D x\,  \Delta ^4(x) .
\label{32}\end{equation}
Invoking once more the  inhomogeneous field equation (\ref{15}) and
Veltman's rule
(\ref{b1}),
 we obtain
the integrals
\begin{equation}
\int d^D x\,  \Delta ^2 _{\mu\mu }(x)  \Delta ^2 (x) =
 - 2 \omega ^2  \Delta^{3} (0) + \omega^4 \int d^D x\,  \Delta ^4 (x),
\label{33}\end{equation}
and
\begin{equation}
\int d^D x\,  \Delta _{\mu\mu }(x)  \Delta ^2_\mu (x)  \Delta (x) =
 \omega^{2} \int d^D x\,  \Delta ^2_\mu (x)  \Delta ^2 (x) .
\label{34}\end{equation}

Due to Eq.~(\ref{32}), the integral
(\ref{34}) takes the form
\begin{eqnarray}
\int d^D x\,  \Delta _{\mu\mu }(x)  \Delta ^2_\mu (x)  \Delta (x) &=&
   \frac{1}{3} \omega^2  \Delta ^3 (0)
 \nonumber \\&-& \frac{1}{3}
  \omega^4 \int d^D x\,  \Delta ^4 (x) .
\label{35}\end{eqnarray}
Partial integration, together with  Eqs.~(\ref{33}) and (\ref{35}),
leads to
\begin{eqnarray}
&&\!\!\!\int d^D x\, \partial_\mu  \Delta _{\lambda\lambda } (x)  \Delta _\mu
  (x)  \Delta ^2 (x) =
 \nonumber \\
 && - \int d^D x\,  \Delta ^2 _{\lambda\lambda }(x)  \Delta ^2 (x)
   - 2\int d^D x\,  \Delta _{\lambda\lambda }(x)  \Delta ^2_\mu (x)  \Delta (x)
 \nonumber \\
 & &= \frac{4}{3} \omega^2  \Delta ^3 (0) - \frac{1}{3}
  \omega^4 \int d^D x\,  \Delta ^4 (x) ,
\label{36}
\end{eqnarray}
A further partial integration, and use of Eqs.~(\ref{28}),
(\ref{34}), and  (\ref{36}),
produces the decompositions
of the second  and third
Feynman diagrams
(\ref{29ff}) and (\ref{29}):
\begin{eqnarray}
\lefteqn{ \!\!\!\!\!\!\!\!\!\!\!
4\,\int d^D x\,  \Delta (x)  \Delta _\mu (x)  \Delta _ \nu (x)
  \Delta _{\mu \nu }(x)  =}\nonumber \label{37}\\
 & = & -2\,I_{D}
+
 4\omega^{2}\int d^D x\,  \Delta ^2 (x)  \Delta _{\mu}^{2} (x) ,
 \label{37}
\end{eqnarray}
and
\begin{eqnarray}
\lefteqn{ \!\!\!\!\!\!\!\!\!\!\!\int d^D x\,  \Delta ^2_\mu (x)  \Delta ^2_ \nu (x)
=}\nonumber \\
& = & I_{D}
 -
3 \omega^{2}\int d^D x\,  \Delta ^2 (x)  \Delta _{\mu}^{2} (x) .
\label{38}
\end{eqnarray}
We now make the important observation that
the subtle integral $I_D$ of Eq.~(\ref{30})
 appears in Eqs.~(\ref{29a}),
(\ref{37}) and (\ref{38})
in such a way that it
drops out
from
the sum of the watermelon diagrams
in (\ref{f5}):
\begin{eqnarray}
&&\quad\quad\quad\hspace{2mm}\raisebox{-2mm}{\mbox{\input 11.tps }}
%\int d^D x\,  \Delta ^2 (x)  \Delta ^2_{\mu \nu }(x)
+
 4\!\!\!\hspace{2mm}\raisebox{-1.9mm}{\mbox{\input 12.tps }} \!~
% \!\!4\int\! d^Dx \Delta (x)
% \Delta _\mu(x) \Delta _ \nu (x) \Delta _{\mu \nu }(x),
+
\!\!\!\hspace{2mm}\raisebox{-2mm}{\mbox{\input 13.tps }}
%\int d^Dx\Delta _\mu(x)
%\Delta _\mu(x)
%\Delta _\nu(x)
%\Delta _\nu(x),
\nonumber\\
&& = \int d^D x\,\Delta ^2 (x)\Delta^2_{\mu\mu }(x)
+  \omega^{2}\int d^D x\,\Delta^2 (x) \Delta _{\mu}^{2} (x).
\label{nss1}
\end{eqnarray}
Using
(\ref{32}) and (\ref{33}),
the right-hand side
becomes
a sum
of completely regular
expressions. Moreover,
adding to this sum
the last two watermelon-like
diagrams in Eq.~(\ref{f5}):
\begin{eqnarray}
 4\omega^2\! \hspace{-27.0mm}\raisebox{-12.3mm}{\mbox{\input waterm2.tps }}
~~~~~~~~~~~
~~~~~~~~~~~
 \!
= 4\omega^2\, \int d^D x\,\Delta^2 (x) \Delta^2_ \mu  (x),
\label{rest17}\end{eqnarray}
~\\[-1.2cm]
and
\begin{eqnarray}
 \! \!\frac{2}{3}\omega^4 \!
\hspace{-27.0mm}\raisebox{-12.3mm}{\mbox{\input waterm.tps }}
~~~~~~~~~~~
~~~~~~~~~~~
=  \frac{2}{3}\omega^4 \,\int d^D x\,
 \Delta ^4 (x) ,
\label{rest18}\end{eqnarray}
~\\[-1.2cm]
we obtain
for the
contribution
of all watermelon-like
diagrams (\ref{f5}) the simple expression
\begin{eqnarray}
&&\!\!\!\!\!\!\!\!\!\!\!\!\quad -F_2^{(4)} =
-2g^2 \,
 \int d^D x\,\Delta^2 (x) \nonumber\\
 &&\times
 \left[\Delta ^2_{\mu\mu} (x) + 5\omega^2\,
 \Delta^2 _\mu (x)
 + \frac{2}{3}\omega^4 \,\Delta ^2 (x)\right]\nonumber\\
&&
\quad\quad =  \frac{2}{3} \,\omega^2 \Delta^3 (0)
\mathop{=}_{D\rightarrow 1}   \frac{1}{12\omega}.
\label{nss2}\end{eqnarray}
This cancels the finite
contribution (\ref{b3}), thus making also the second-order
free energy in (\ref{m7}) vanish, and
confirming the invariance of the perturbatively defined
path integral under coordinate transformations up to this order.

\section{Summary}
In this note we have set up
simple  rules for calculating
integrals over products of distributions in configuration space
which produce the same results
as
dimensional regularization
in momentum space.
For a path integral of a quantum-mechanical point particle
in a harmonic potential, we have shown that these rules lead
to a reparametrization-invariant
perturbation expansions of path integral.

Let us end with the remark
that
in
the
time-sliced
definition
of path integrals,
reparametrization invariance
has been established
as long time ago
in the textbook
\cite{4}.

%Noninvariance under reparametrizations
%were a major source of problems
%in perturbative as well as
%and time-sliced definitions of path integrals in Ref.~\cite{4}.

%

\end{document}